\documentclass[sigchi-a, authorversion]{acmart}
\usepackage{booktabs} 
\usepackage{ccicons}  
\usepackage{listings}
\usepackage{smartdiagram}
\usepackage{listings}

\newcommand{\Cag}[0]{Conversational agent}
\newcommand{\cag}[0]{conversational agent}
\newcommand{\bug}[0]{breakdown}
\newcommand{\Fix}[1]{\textbf{Amin:#1}}

\newcommand{\Part}[1]{\noindent\textbf{\emph{#1}}}

\acmConference[]{Work in progress}

\begin{document}
\title{An Automated Testing Framework for Conversational Agents}

\author{Soodeh Atefi}
\affiliation{%
  \institution{Department of Computer Science}
  \institution{University of Houston}
  \city{Houston}
  \state{TX}
  \country{USA} }
\email{satefi@uh.edu}

\author{Mohammad Amin Alipour}
\affiliation{%
  \institution{Department of Computer Science}
  \institution{University of Houston}
  \city{Houston}
  \state{TX}
  \country{USA}}
\email{alipour@cs.uh.edu}

%
%



\begin{abstract}
\Cag{}s are systems with a conversational interface that afford interaction in spoken language.
These systems  are becoming prevalent and are  preferred in various contexts and for many users.
Despite their increasing success, the automated testing infrastructure to support the effective and efficient development of such systems compared to traditional software systems is still limited.

Automated testing framework for conversational systems can improve the quality of these systems by assisting developers to write, execute, and maintain test cases. 
In this paper, we introduce our work-in-progress automated testing framework, and its realization in the Python programming language. 
We discuss some research problems in the development of such an automated  testing framework for \cag{}s.
In particular, we point out the problems of the specification of the expected behavior, known as test oracles, and semantic comparison of utterances. 
\end{abstract}

\maketitle
\section{Introduction}
\Cag{}s are systems with a conversational interface that afford interaction in spoken language.
A \cag{} reasons about user's intent from the utterances and acts upon it.
More sophisticated agents may include extraneous factors such as intonation or conversation environments in reasoning about the user's intent and its expected behavior.

%
Conversational agents can assist or replace humans in hostile or highly sensitive environments.
For example, a \cag{} can help in understaffed operation rooms or rescue operations. 
In such environments, \cag{}s can significantly improve user's effectiveness and agility by reducing or removing the need for looking at a display or communicating with the system via keyboards, mouse, or other physical interaction devices. 

Unlike traditional software systems, the evaluation of  \cag{}s is highly ad-hoc, heavily empirical and involves human subjects to participate in user studies to interact with a \cag.
Moreover, there is no tooling to support automated testing of \cag{}s.
Currently, testing a conversational system constitutes writing the test requests to and recording responses from an agent in various files with different audio and text formats.
Lack of reliable development infrastructure and environments would impact a developer's experience and impede the development and maintenance of conversational systems.

We set out to build a software infrastructure for effective and efficient development of conversational agents. 
Our goal is to improve the developer's experience in developing reliable conversational agents. 
In this paper, we report our progress in building one component of this project: an automated testing framework for conversational agents. 
Our vision for an automated testing framework is to build a system to allow developers to write, execute, and maintain test cases for these agents. 
We aim to provide an abstraction at the programming language level for testing.
Ideally, this abstraction should hide the unnecessary details of file system dependency, sound generation and acquisition, and natural language comparisons.
The goal of automated testing is to facilitate finding bugs by exposing failures in systems. 
Therefore, we briefly discuss the notion of failure in conversational agents. 

\Part{Paper organization.}
In the rest of this paper, we first describe the notion of failure in conversational systems. 
We then introduce our current prototype for automated testing. 
Next, we discuss the related work in user simulation and graphical user interface testing.
Finally, we conclude the paper with a discussion on some of the problems that we are attempting to solve through our ongoing efforts.

\section{Failure in \cag{}}
\label{sec:example}
In this section, we discuss the notion of failure in conversational agents. 
In the dialog community a failure in conversational systems is also referred to as conversation breakdown. 

\begin{marginfigure}
    \includegraphics[width=\marginparwidth]{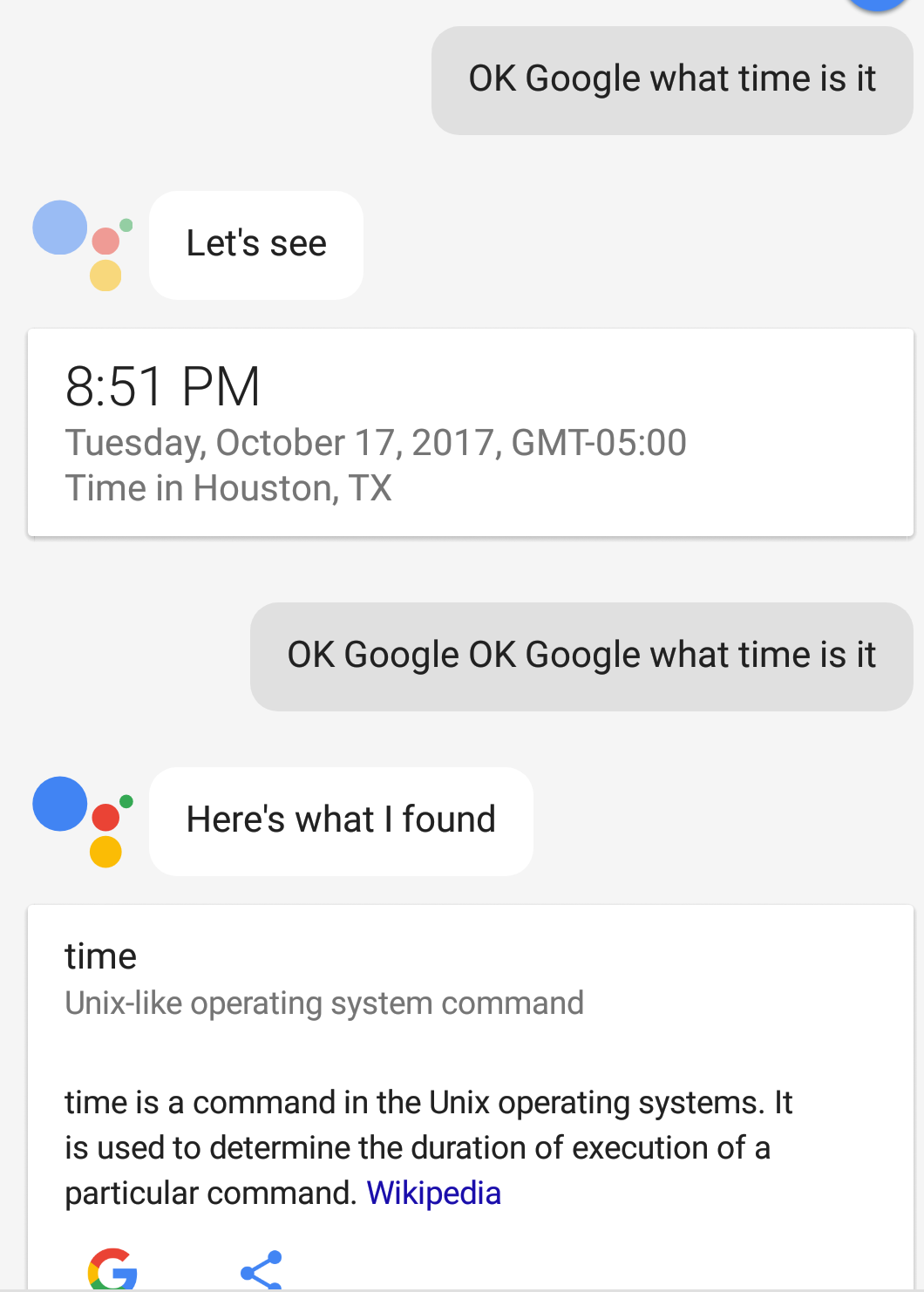}
    \caption{An example of bug in the Google Assistant system}
    \label{fig:bug}
\end{marginfigure}

A Breakdown is a situation in which a conversational agent gives an incoherent answer and the user cannot proceed with the conversation~\cite{Higashinaka2016}.
There are three categories of breakdowns in chat-oriented dialog agents which can be considered a breakdown. In the first category, the system's answer is not understandable. In other words, the answer is not relevant to the context of the conversation. For example, the user says, "it's hot today, isn't it?" and the system answers "Please tell me your favorite movie genre". In this example, the system's response has no relation to the user's utterance. In the second category, the system is unable to answer the user's question and ignores the question. For instance, the user asks about the weather and the system answers by showing some links related to the keywords of the user's utterance. In the third category, the system's answer has unclear intention (that is, the purpose of the system's utterance is ambiguous). For example, the user asks "Do you know what movie will be aired on Friday night?" and the system answers "Yes, yes"~\cite{Higashinaka2015b}. 

Figure~\ref{fig:bug} illustrates a breakdown, in Google's conversational agent, Google Assistant, released in Summer 2017 on Android 8.0.
Google Assistant expects users to start a conversation with ``OK Google'' utterance. Figure~\ref{fig:bug} shows that in response to ``OK Google, what time is it?'', the agent, rightly, returns the time. 
However, the agent's response to a very similar question but with a small perturbation, i.e. duplicate ``OK Google``, is wildly different; the response is a recommendation for a mobile app.
Clearly, the user's intent in both interactions is the same: knowing the time.
Both questions are similar except in the latter dialog ``OK Google'' is uttered twice, perhaps due to the user's mistake, but the agent's response is unreasonably different.

\section{ Testing Framework for \cag{}}
\label{sec:testing}
In this section, we explain our current automated testing framework for conversational agents.

Effective test cases have three main components:
(1) test input, (2) test oracles, and (3) test environment. 
\emph{Test input} is the actual input to the system. 
In conversational agents, a test input is utterances that mimic the user input  to the conversational agent. 
\emph{Test oracles} essentially decide whether the system's behavior adheres to the expected behavior specified by developers in response to the test input. For example, they can check whether the task at hand is actually finished by the agent, or the agent's response is an appropriate response to the user's request. 
\emph{Test environment} specifies the context where the interaction with the system takes place. For example, in conversational agents, it can include the lexicons used, the speed of conversation, or the ambient noise.

We implemented a prototype that allows writing, executing, and maintaining test cases for conversational agents. 
Our framework constitutes three main components: (1)  semantic evaluation, (2)  context builder, and (3) developer interface. 

\emph{Semantic evaluation component} realizes test oracles for the domain of conversational agents. 
The main task of semantic evaluation module is to perform semantic analysis and compare the utterances.
For example, it can decide whether the results of the conversational agents have expected meaning, or whether the generated utterance by the agent is polite. 
Currently, we implemented a semantic comparison technique that is based on the average of the word2vec embedding~\cite{AverageWord2Vec} of the terms used in utterances to approximate their semantic similarity.
This module is extensible to accommodate other similarity techniques such as Google's Universal Sentence Encoder~\cite{GoogleSentEncoder} and Facebook's InferSent sentence embeddings~\cite{FacebookInferSent}.

\emph{Context builder component} uses a builder design pattern~\cite{designpatterns} to bind values to various dialog parameters, e.g., word per second, intonations, whether to allow confirmations, and a threshold for equivalence.
Context builder component also determines the datasets needed for the functionality of the semantic evaluation module.

\emph{Developer interface module} provides a programming interface for developers to use the framework.
Currently, we have implemented  \texttt{assert\_equivalent} which compares the response of the conversational agent with the expected response.
Our framework supports asserting predicates about any visible part of the agent's state, e.g., completion of tasks, or intents recognized by the agent. 
The assertion can also be about the state of the program that uses the conversational interface; for example, in a calendar agent, a predicate can be whether the expected meeting has been scheduled and added to the agenda.

\subsection{Example}

Our framework has been implemented as a \texttt{DiagTest} class that extends Python's \texttt{unittest} framework. 
The Snippet in~\autoref{code:unittest} depicts two test cases \texttt{test\_simple} and \texttt{test\_complex} in our framework.

\texttt{test\_simple} demonstrates a semantic comparison of ``hi'' and ``hello'' by specifying Google News word2vec~\cite{GoogleNewW2V} embeddings as the basis of the comparison. By default, the threshold for the semantic similarity of two utterances is 0.5. However, it can be modified in the context builder component.

\texttt{test\_complex} demonstrates an interaction with a clock agent \texttt{"ca"} which checks the response from the agent.
The first line in this method instructs the framework to use GloVe word representations~\cite{GloVe} in the comparison.

The current implementation only accepts text, however, in the future, we plan to add support for using sound files as input. 
Class \texttt{"Utterance"} implements the basic operations for normalizing the utterances needed for natural language operations. It also transforms an utterance to its vector space encoding. 


\begin{lstlisting}[language=Python,caption={An example of Unit Test for conversational agent},label={code:unittest}, basicstyle=\footnotesize]
class TestDemo(DiagTest):
    def test_simple(self):
        Utterance.model = DiagTest.word2vec
        self.assert_equivalent(Utterance('hello'), Utterance('hi'), "Basic greeting test failure")

    def test_complex(self):
        Utterance.model = DiagTest.glove
        response = ca.getresponse('alarm for six a.m.')
        self.assert_equivalent(Utterance(response), Utterance("You're alarm set of six a.m."))
\end{lstlisting}

\section{Related Work}
\label{sec:related}
The related work can be classified into two broad groups.
The first group includes techniques that attempt to reduce the user studies in the evaluation of conversational agents by building models that simulate human behavior in interactions with conversational agents. 
In the dialog community, there have been efforts to automatically detect dialog breakdowns~\cite{Higashinaka2016}, and classify breakdowns~\cite{Higashinaka2015}.

The second group includes techniques and tools related to testing traditional user interfaces, i.e. graphical user interfaces (GUI). 
Automated testing of graphical user interfaces has been extensively studied~\cite{guisurvey}.
Over the years, several test adequacy metrics for these interfaces have been proposed~\cite{memon2001coverage}.
There are also several tools and techniques for analyzing graphical user interfaces and running test cases to  satisfy test requirements~\cite{brooks2007automated}.

\section{Discussion and Concluding Remarks}
\label{sec:discussion}
Conversational agents have widespread applications in many domains, and can vastly enhance the user experience.
In this paper, we described a prototype to support automated testing of conversational agents. The overarching goal of our project in general, and this framework in particular, is to improve the developers' experience, and improve the quality of conversational agents. 

We believe there are two main threats to creating \emph{fully} automatic testing frameworks for conversational agents: (1) formalizing a fluent conversation, and (2)  test oracles for natural language.

Natural conversation is a human phenomenon that misunderstanding, clarification, and confirmation are inseparable parts of it. 
Therefore, it is inherently difficult to formalize an acceptable behavior of a conversational agent to objectively gauge the fluency of a conversation, and detect breakdowns.
Liu et al. observed that several metrics in the evaluation of dialog systems do not strongly correlate with humans ~\cite{Liu2017}.
As the field evolves, we hope that new computational insights into fluency of conversations would enable us to better understand different aspects of a fluent conversation and provide tools to evaluate conversational agents objectively.

Semantic analysis of natural language is non-trivial. 
Test predicates that involve extracting and comparing the meaning of the utterances would be hard to realize and maintain. 
However, techniques and models for semantic analysis of natural language are continually improving.
We hope as the filed progresses, we can use more reliable models for our framework.

Our framework is still in the early stages of development and needs to be evaluated by real users, i.e. developers.
After integrating more features, we plan to make the framework available to developers and practitioners to evaluate its usefulness, and its impact on the development life-cycle of conversational agents. 
As for new features, we plan to extend our framework to allow audio files and text-to-speech modules to interact with the agent.
We are also developing an automata-based test generation technique for VoiceXML-based~\cite{voicexml} systems to automatically generate test inputs.

\bibliographystyle{acm}
\bibliography{bib}

\begin{thebibliography}{10}

\bibitem{guisurvey}
{\sc Banerjee, I., Nguyen, B., Garousi, V., and Memon, A.}
\newblock Graphical user interface (gui) testing: Systematic mapping and
  repository.
\newblock {\em Information and Software Technology 55}, 10 (2013), 1679--1694.

\bibitem{brooks2007automated}
{\sc Brooks, P.~A., and Memon, A.~M.}
\newblock Automated gui testing guided by usage profiles.
\newblock In {\em Proceedings of the twenty-second IEEE/ACM international
  conference on Automated software engineering\/} (2007), ACM, pp.~333--342.

\bibitem{GoogleSentEncoder}
{\sc Cer, D., Yang, Y., Kong, S.-y., Hua, N., Limtiaco, N., John, R.~S.,
  Constant, N., Guajardo-Cespedes, M., Yuan, S., Tar, C., et~al.}
\newblock Universal sentence encoder.
\newblock {\em arXiv preprint arXiv:1803.11175\/} (2018).

\bibitem{FacebookInferSent}
{\sc Conneau, A., Kiela, D., Schwenk, H., Barrault, L., and Bordes, A.}
\newblock Supervised learning of universal sentence representations from
  natural language inference data.
\newblock In {\em Proceedings of the 2017 Conference on Empirical Methods in
  Natural Language Processing\/} (2017), pp.~670--680.

\bibitem{designpatterns}
{\sc Gamma, E., Helm, R., Johnson, R., and Vlissides, J.}
\newblock {\em Design Patterns: Elements of Reusable Object-oriented Software}.
\newblock Addison-Wesley Longman Publishing Co., Inc., Boston, MA, USA, 1995.

\bibitem{Higashinaka2015}
{\sc Higashinaka, R., Funakoshi, K., and Araki, M.}
\newblock {Towards Taxonomy of Errors in Chat-oriented Dialogue Systems}.
\newblock In {\em Sigdd2015\/} (2015), no.~September, pp.~87--95.

\bibitem{Higashinaka2016}
{\sc Higashinaka, R., Funakoshi, K., Kobayashi., Y., and Inaba, M.}
\newblock {The dialogue breakdown detection challenge: Task description,
  datasets, and evaluation metrics}.
\newblock {\em 10th edition of the Language Resources and Evaluation
  Conference\/} (2016), 3146--3150.

\bibitem{Higashinaka2015b}
{\sc Higashinaka, R., Mizukami, M., Funakoshi, K., Araki, M., Tsukahara, H.,
  and Kobayashi, Y.}
\newblock {Fatal or not? Finding errors that lead to dialogue breakdowns in
  chat-oriented dialogue systems}.
\newblock In {\em EMNLP\/} (2015), pp.~2243--2248.

\bibitem{Liu2017}
{\sc Liu, C.-W., Lowe, R., Serban, I., Noseworthy, M., Charlin, L., and Pineau,
  J.}
\newblock How not to evaluate your dialogue system: An empirical study of
  unsupervised evaluation metrics for dialogue response generation.
\newblock In {\em Proceedings of the 2016 Conference on Empirical Methods in
  Natural Language Processing\/} (2016), pp.~2122--2132.

\bibitem{voicexml}
{\sc Lucas, B.}
\newblock Voicexml for web-based distributed conversational applications.
\newblock {\em Communications of the ACM 43}, 9 (2000), 53--57.

\bibitem{memon2001coverage}
{\sc Memon, A.~M., Soffa, M.~L., and Pollack, M.~E.}
\newblock Coverage criteria for gui testing.
\newblock {\em ACM SIGSOFT Software Engineering Notes 26}, 5 (2001), 256--267.

\bibitem{GoogleNewW2V}
{\sc Mikolov, T., Chen, K., Corrado, G., and Dean, J.}
\newblock Efficient estimation of word representations in vector space.
\newblock {\em ICLR\/} (2013).

\bibitem{AverageWord2Vec}
{\sc Mitchell, J., and Lapata, M.}
\newblock Vector-based models of semantic composition.
\newblock {\em Proceedings of ACL\/} (2008), 236--244.

\bibitem{GloVe}
{\sc Pennington, J., Socher, R., and Manning, C.}
\newblock Glove: Global vectors for word representation.
\newblock In {\em EMNLP\/} (2014), pp.~1532--1543.

\end{thebibliography}

\end{document}